\PassOptionsToPackage{table}{xcolor}

\newcommand{\heatblue}[2]{%
  \begingroup
  \setlength{\fboxsep}{1pt}%
  \colorbox{blue!#1}{\strut #2}%
  \endgroup
}

\newcommand{\heatorange}[2]{%
  \begingroup
  \setlength{\fboxsep}{1pt}%
  \colorbox{orange!#1}{\strut #2}%
  \endgroup
}
\documentclass[cameraready]{Interspeech}
\usepackage{booktabs}
\usepackage{xcolor}
\usepackage{url}
\usepackage{colortbl}
\usepackage[table]{xcolor}
\usepackage{colortbl}
\usepackage{comment}
\usepackage[table]{xcolor}   
\usepackage{colortbl}
\usepackage{pgf}             

\title{DiffAnon: Diffusion-based Prosody Control for Voice Anonymization}

\author[affiliation={1}, orcid=0000-0003-4593-9057]{Ismail Rasim}{Ulgen}
\author[affiliation={1}, orcid=0009-0007-2658-2220]{Zexin}{Cai}
\author[affiliation={2}, orcid=0000-0002-6097-9164]{Nicholas} {Andrews}
\author[affiliation={1}, orcid=0000-0003-1565-064X]{Philipp}{Koehn}
\author[affiliation={1}, orcid=0000-0001-8078-3305]{Berrak}{Sisman}

\address{
 $^1$ Center for Language and Speech Processing, Johns Hopkins University, USA \\
      $^2$ Human Language Technology Center of Excellence (COE), Johns Hopkins University, USA
}

\email{iulgen1@jhu.edu, noa@jhu.edu, phi@jhu.edu, sisman@jhu.edu}


\newcommand\blfootnote[1]{%
  \begingroup
  \renewcommand\thefootnote{}\footnote{#1}%
  \addtocounter{footnote}{-1}%
  \endgroup
}

\usepackage{comment}
\usepackage{cite}
\usepackage{highlight}
\usepackage{colortbl}
\usepackage{amsmath}
\usepackage{multirow}

\begin{document}
\maketitle
\begin{abstract}
To preserve or not to preserve prosody is a central question in voice anonymization. Prosody conveys meaning and affect, yet is tightly coupled with speaker identity. Existing methods either discard prosody for privacy or lack a principled mechanism to control the utility–privacy trade-off, operating at fixed design points. We propose DiffAnon, a diffusion-based anonymization method with classifier-free guidance (CFG) that provides explicit, continuous inference-time control over prosody preservation. DiffAnon refines acoustic detail over semantic embeddings of an RVQ codec, enabling smooth interpolation between anonymization strength and prosodic fidelity within a single model. To the best of our knowledge, it is the first voice anonymization framework to provide structured, interpolatable inference-time prosody control. Experiments demonstrate structured trade-off behavior, achieving strong utility while maintaining competitive privacy across controllable operating points.

\end{abstract}

\section{Introduction}

Voice anonymization aims to protect speaker privacy by concealing speaker identity in a speech signal, while still conveying the intended linguistic content for effective communication \cite{TOMASHENKO2022101362}. Speech jointly encodes linguistic content, para-linguistic information, and speaker identity, all of which are inherently entangled. As a result, voice anonymization faces a fundamental challenge: enhancing privacy often degrades utility, whereas preserving utility may reveal speaker identity, leading to an inherent trade-off between utility and privacy~\cite{zhang2023voicepm, 10832351}. 


A key factor shaping the utility–privacy trade-off in voice anonymization is prosody, which conveys emphasis, emotion, and speaker intent \cite{prosody_meaning}. For example, the utterance \textit{“You did that“} can express a neutral statement, an accusation, disappointment, or a question depending on its prosodic realization. Preserving prosody is therefore essential for maintaining meaning and expressiveness \cite{Cutler1997-fc,Gussenhoven_2004}. At the same time, prosody is closely linked to speaker identity, as speakers exhibit characteristic prosodic patterns \cite{helander07_interspeech,pro_spk,Mary2012}. Modern speaker recognition systems can exploit such prosodic cues, making prosody a potential source of identity leakage \cite{carbonneau25_ssw,rasim}. As a result, prosody lies at the center of the utility–privacy trade-off in voice anonymization: preserving it improves utility but can compromise privacy, while modifying or suppressing it strengthens anonymization at the cost of degraded expressiveness and meaning \cite{tomashenko2026voiceprivacychallengepreservingemotional,vpc2022,10832351}.

Existing voice anonymization approaches adopt different strategies to navigate the utility–privacy trade-off. Cascaded ASR–TTS systems~\cite{meyer22_spsc,panariello24_spsc } prioritize privacy by transcribing speech into text and resynthesizing it, effectively removing speaker-specific acoustic characteristics while discarding the source prosody, which reduces expressiveness and para-linguistic fidelity. In contrast, voice conversion (VC)–based anonymization methods~\cite{fang19_ssw_vcbased,pierre22_interspeech_vcbased,akti24_spsc,diffvcplus_interspeech} aim to preserve linguistic content and aspects of source prosody while modifying speaker identity, but imperfect disentanglement can lead to residual identity leakage. Some approaches additionally apply random perturbations to prosodic features~\cite{prosody_cloning_perturb,Singh2024,franzreb2025privateknnvcinterpretableanonymization}, obscuring identity at the cost of unpredictable utility degradation. Across these methods, prosody handling is largely fixed by design, with systems suppressing, implicitly preserving, or heuristically perturbing it. While recent work has explored combining multiple anonymization systems for flexibility~\cite{henry}, they do not enable continuous, inference-time control within a single model. Consequently, most existing systems operate at largely fixed points on the utility–privacy trade-off without explicit prosody control.

To address these limitations, we propose \textbf{DiffAnon}, a diffusion-based voice anonymization framework with classifier-free guidance (CFG) \cite{ho2021classifierfree} that is specifically designed to enable controllable anonymization. CFG brings adjustable guidance during anonymization of speech without the need of external classifiers which are lacking for prosody. DiffAnon introduces a systematic mechanism for modifying source prosody by adjusting the strength of prosodic conditioning during inference, thereby controlling how much source prosody is preserved in anonymized speech. It formulates anonymization as an iterative refinement of acoustic details over semantic representations obtained from a residual vector quantization (RVQ)–based speech codec~\cite{zhang2024speechtokenizer}. By refining acoustic information on top of speaker-agnostic semantic embeddings, the diffusion process aligns naturally with the iterative refinement structure of RVQ, making it well suited for controlled anonymization. This formulation allows CFG to directly regulate the contribution of source prosody during generation, enabling continuous and interpretable adjustment of the utility--privacy trade-off within a single trained model. The structured generation formulation of diffusion allows DiffAnon to retain meaningful speech utility while flexibly adjusting anonymization strength. \blfootnote{\textbf{DiffAnon is available at:} \scriptsize \url{https://github.com/rsmlgen/diffanon}}

Our contributions can be summarized as: (i) we introduce DiffAnon, the first voice anonymization framework that provides explicit control over source prosody preservation via classifier-free guidance; (ii) we propose a novel diffusion-based formulation that refines acoustic information over speaker-agnostic semantic embeddings in an RVQ-based codec space; and (iii) we show on VoicePrivacy Challenge 2024 that DiffAnon achieves strong utility and competitive privacy with systematic utility–privacy trade-off navigation. We publicly release the codes and pretrained models to enable reproducibility.



\begin{figure*}[!t]
\vspace{-4mm}
    \centering
    \scalebox{0.28}
    {\includegraphics{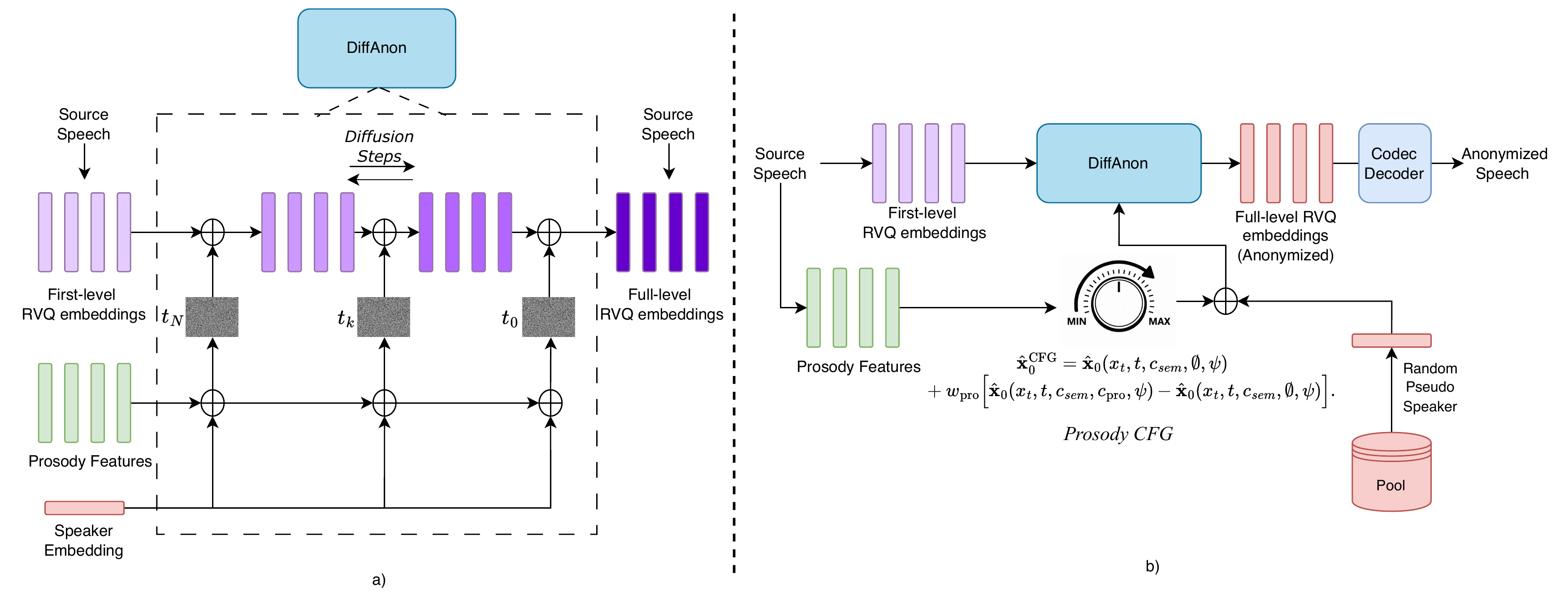}}
    \vspace{-4mm}
    \caption{a) Conditional diffusion training to construct codec embeddings b) Anonymization inference via adjusted prosody with CFG and pseudo-speaker condition (randomly sampled from a pseudo-speaker pool)} 
    
    \label{fig:framework}

\vspace{-5mm}
\end{figure*}

\section{The Utility--Privacy Trade-off}
Voice anonymization is inherently a multi-objective problem, as speech simultaneously encodes linguistic content, speaker identity, prosody, and emotion. As a result, anonymization must be evaluated along competing dimensions rather than a single metric. The VoicePrivacy initiative~\cite{TOMASHENKO2022101362,vpc2022,tomashenko2026voiceprivacychallengepreservingemotional} formalizes this utility--privacy trade-off through standardized evaluation protocols. Privacy is measured as resistance to speaker verification attacks under different attacker knowledge assumptions, typically quantified by equal error rate (EER)~\cite{KINNUNEN201012}. Utility is assessed through linguistic preservation using word error rate (WER), and para-linguistic preservation using task-specific metrics such as emotion recognition recall (UAR) in VoicePrivacy 2024~\cite{tomashenko2026voiceprivacychallengepreservingemotional} and F0 correlation in VoicePrivacy 2022~\cite{vpc2022}. Across challenge editions, results consistently show that stronger anonymization often degrades prosodic and emotional fidelity, while better preservation of para-linguistic information increases the risk of speaker re-identification~\cite{vpc2022,zhang2023voicepm, tomashenko2026voiceprivacychallengepreservingemotional,10832351}. Consequently, most existing approaches operate at mostly fixed design points on this trade-off, lacking mechanisms for continuous, inference-time adjustment. In contrast, our approach enables explicit and controllable navigation of the utility--privacy trade-off through prosody control within a single model.

\section{DiffAnon}
We introduce DiffAnon, a diffusion-based framework for controllable voice anonymization (Fig.~\ref{fig:framework}). The model is trained to reconstruct SpeechTokenizer codec embeddings~\cite{zhang2024speechtokenizer} conditioned on separately extracted representations of linguistic content, prosody, and speaker identity, which are then decoded into waveform via the codec decoder. At inference, DiffAnon anonymizes speech by preserving source content and prosody while replacing the original speaker identity with a pseudo-speaker embedding. The degree of prosody preservation is regulated through classifier-free guidance (CFG), enabling explicit and continuous control of anonymization strength.




\textbf{Content Condition}. As the content condition, DiffAnon utilizes the first-level quantization embeddings of SpeechTokenizer~\cite{zhang2024speechtokenizer}, an RVQ-based neural speech codec. SpeechTokenizer encodes speech into multi-level tokens $Q^{1:8}$ using residual vector quantization and applies semantic distillation to the first-level tokens $Q^1$. Prior work shows that $Q^1$ primarily captures linguistic content while containing minimal speaker information. In SpeechTokenizer, higher-level tokens $Q^{2:8}$ refine acoustic details on top of the semantic tokens $Q^1$ within an interpolatable embedding space. DiffAnon is inspired by this formulation and proposes a new diffusion paradigm tailored for voice anonymization. It treats $Q^1$ as a semantic prior and predicts the full codec embeddings $\hat{Q}^{1:8}$ by refining the remaining acoustic components corresponding to subsequent RVQ stages. 

\textbf{Prosody Condition.} We model prosody using frame-level latent features $z_{mpm}$ extracted from the masked prosody model (MPM)~\cite{wallbridge25_interspeech}. MPM is trained via masked prediction of speaker-normalized pitch, energy, and voiced/unvoiced indicators, enabling it to capture contextual prosodic structure while minimizing linguistic and speaker identity information. This makes $z_{mpm}$ a suitable representation for controllable prosody preservation.

\textbf{Speaker Condition.} Speaker identity is represented using the pre-trained FreeVC speaker encoder~\cite{freevc}, which produces an utterance-level embedding $z_{spk}$. Using a single utterance-level vector limits leakage of frame-level prosodic cues into the speaker representation and simplifies pseudo-speaker sampling for anonymization. This representation provides stable and explicit speaker conditioning during generation.

\vspace{-3mm}
\subsection{Conditional Diffusion Formulation}

We adopt a denoising diffusion probabilistic model (DDPM) \cite{ddpm} formulation. Noise is gradually added to target embeddings $\mathbf{x}_0$ over $T$ timesteps:
\setlength{\abovedisplayskip}{4pt}
\setlength{\belowdisplayskip}{5pt}
\begin{equation}
\mathbf{x}_t = \sqrt{\bar{\alpha}_t}\,\mathbf{x}_0
+ \sqrt{1-\bar{\alpha}_t}\,\boldsymbol{\epsilon},  \quad \boldsymbol{\epsilon} \sim \mathcal{N}(\mathbf{0}, \mathbf{I}).
\end{equation}
where $\beta_t$ defines the variance schedule of the forward diffusion process, $\alpha_t$ and $\bar{\alpha_t}$ are derived from $\beta_t$ as
\setlength{\abovedisplayskip}{0pt}
\setlength{\belowdisplayskip}{3pt}
\begin{equation}
        \alpha_t = 1 - \beta_t, \quad
    \bar{\alpha}_t = \prod_{s=1}^{t} \alpha_s
\end{equation}
DiffAnon uses $x$-prediction, directly predicting the clean embeddings $\mathbf{\hat{x}}_0$ at timestep $t$ given noisy input $\mathbf{x}_t$ and conditions $c_{sem}, c_{pro}, c_{spk}$ which are semantic, prosody and speaker conditions and trained with the following loss:
\setlength{\abovedisplayskip}{4pt}
\setlength{\belowdisplayskip}{4pt}
\begin{equation}
\mathcal{L}_{x_0}
=
\mathbb{E}
\left[
||
\mathbf{x}_0 -
\hat{\mathbf{x}}_{0,\theta}(\mathbf{x}_t, t, c_{sem}, c_{pro}, c_{spk})
||^2
\right].
\end{equation}

In this formulation, the target $\mathbf{x}_0$ corresponds to the full SpeechTokenizer embeddings after all quantization stages, $Q^{1:8} \in \mathbb{R}^{1024 \times T}$, where $1024$ denotes the embedding dimension and $T$ the number of frames. The content condition $c_{sem}$ is given by the first-level SpeechTokenizer embeddings $Q^1 \in \mathbb{R}^{1024 \times T}$. The prosody condition $c_{pro}$ consists of frame-level MPM features $z'_{mpm} \in \mathbb{R}^{256 \times T}$, which are linearly interpolated to match the frame rate of $Q^{1:8}$. The speaker condition $c_{spk}$ is the utterance-level embedding from the FreeVC speaker encoder, $z'_{spk} \in \mathbb{R}^{256 \times T}$, obtained by repeating the 256-dimensional vector across $T$ frames to align with the other features. We utilize conditioning by addition, more specifically, adding the projected conditions to the noisy input $\mathbf{x}_t$. Specifically, $c_{pro}$ and $c_{spk}$ are passed through separate convolutional projection modules, $proj_{pro}$ and $proj_{spk}$, projected to the same dimension before being added to the latent representation.


The key distinctive aspect of our proposed formulation for anonymization lies in content conditioning. We directly add the content condition $c_{sem}$ at each timestep without any projection. Since $Q^1$ is already provided to the model, this encourages the model to focus on predicting the remaining acoustic components $Q^{2:8}$ when reconstructing $Q^{1:8}$. This simplifies the task and enables more effective control in conditional generation with respect to prosody and speaker, as the model does not need to predict the full semantic–acoustic representation from scratch. Moreover, conditioning on $Q^1$ at every timestep ensures a base level of content preservation.

\subsection{CFG for Voice Anonymization}

We employ classifier-free guidance (CFG)~\cite{ho2021classifierfree} during inference to control the strength of source prosody conditioning. Our goal in anonymization is to replace the source speaker identity with a pseudo-speaker $\psi$ that is not identifiable with the original speaker. CFG enables adjustable conditioning by combining conditional and unconditional predictions. In our setting, “unconditional” refers to dropping a specific condition by using null as the conditioning input. We consider two guidance modes: (i) prosody-adjusted guidance and (ii) pseudo-speaker guidance.


\textbf{Prosody-adjusted guidance.} In this setting, we aim to keep different degrees of source prosody for utility. For prosody adjustment we have the following CFG formulation
\refstepcounter{equation}
{\footnotesize
\begin{flalign}
&\hat{\mathbf{x}}_{0,\theta}^{\mathrm{CFG}}
=\hat{\mathbf{x}}_{0,\theta}(x_t,t,c_{sem}, \emptyset,\psi) \tag{\theequation}\\
&\quad + w_{\mathrm{pro}}
\Big[
\hat{\mathbf{x}}_{0,\theta}(x_t,t,c_{sem},c_{\mathrm{pro}},\psi)
-
\hat{\mathbf{x}}_{0,\theta}(x_t,t,c_{sem},\emptyset,\psi)
\Big]. \notag &&
\end{flalign}}

\begin{table*}[!h]
\vspace{-5mm}
\centering
\scriptsize
\caption{Anonymization results (utility vs privacy) under the VoicePrivacy Challenge 2024 protocol. Higher EER indicates stronger privacy. Lower WER and higher UAR/F0 indicate better utility. “Prosody” denotes the prosody guidance weight $w_{\mathrm{pro}}$. Shaded cells highlight DiffAnon trends across prosody control settings (darker indicates better according to arrows column-wise).}
\vspace{-3mm}
\setlength{\tabcolsep}{3pt}
\begin{tabular}{l c c @{\hspace{1.5pt}}c c c c| c c| c c| c @{\hspace{1pt}}c}
\toprule
\textbf{Method}  & \textbf{Prosody} & \textbf{Speaker} &
\multicolumn{4}{c|}{\textbf{Privacy (EER, \%) $\uparrow$}} &
\multicolumn{2}{c|}{\textbf{Utility (WER, \%) $\downarrow$}} &
\multicolumn{2}{c|}{\textbf{Utility (UAR, \%) $\uparrow$}} &
\multicolumn{2}{c}{\textbf{Utility (f0-corr, \%) $\uparrow$}} \\
\cline{4-7}\cline{8-9}\cline{10-11}\cline{12-13}
 & & &
\multicolumn{2}{c}{\textit{dev}} & \multicolumn{2}{c|}{\textit{test}} &
\textit{dev} & \textit{test} &
\textit{dev} & \textit{test} &
\textit{dev} & \textit{test} \\
 & & & lazy & semi & lazy & semi & - & - & - & - & - & - \\
\midrule
Ground Truth & --  & -- & - & - & - & - & 1.80 & 1.84 & 69.07 & 71.06 & -- & -- \\
\midrule
B1* & --  & pseudo & -- & 9.20 & -- & 6.07 & 3.07 & 2.91 & 42.71 & 42.78 & -- & -- \\
B2 & --  & pseudo & 31.52 & 8.45 & 29.99 & 4.48  & 10.48 & 9.99 & \textbf{55.64} & \textbf{53.49} & 58.70 & 56.82 \\
B3 &  -- & pseudo & 44.08 & 25.74 & 44.73 & 27.38 & 4.16 & 4.29 & 37.86 & 36.27 & 37.49 & 36.58 \\
B4 &  -- & pseudo & \textbf{50.14} & 32.89 & 48.84 & 30.59 & 6.12 & 5.90 & 42.19 & 42.00 & 70.93 & 69.39 \\
B5 & -- & pseudo & 48.68 & 34.37 & \textbf{49.55} & 34.34 & 4.81 & 4.31 & 38.95 & 37.45 & 66.64 & 65.67 \\
B6* & -- & pseudo & -- & 23.05 & -- & 21.14 & 9.69 & 9.09 & 36.09 & 36.13 & -- & -- \\
\midrule 
T8* \cite{henry} & -- & -- & -- & 40.93 & -- & 40.70 & 3.45 & 3.19 & 47.07 & 47.10 & 49.22 & -- \\
T9* \cite{t9} & -- & -- & -- & 33.43 & -- & 35.10 & \textbf{2.33} & \textbf{2.37} & 60.09 & \textbf{60.95} & -- & -- \\
T10* \cite{t10} & -- & -- & -- & \textbf{42.45} & -- & \textbf{42.46} & 3.51 & 3.19 & \textbf{62.93} & 60.87 & -- & -- \\
\midrule 

DiffAnon   & 1.0 & $\psi$ &
\heatblue{10}{35.23} & \heatblue{10}{14.97} &
\heatblue{10}{33.09} & \heatblue{10}{14.53} &
4.91 & 4.62 &
\heatorange{68}{52.32} & \heatorange{70}{50.80} &
\heatorange{70}{\textbf{76.67}} & \heatorange{70}{\textbf{75.58}} \\

DiffAnon   & 0.8 & $\psi$ &
\heatblue{21}{37.74} & \heatblue{14}{15.73} &
\heatblue{15}{34.40} & \heatblue{15}{15.28} &
5.11 & 4.75 &
\heatorange{64}{51.58} & \heatorange{67}{50.38} &
\heatorange{64}{74.70} & \heatorange{65}{73.82} \\

DiffAnon   & 0.5 & $\psi$ &
\heatblue{40}{41.74} & \heatblue{32}{19.57} &
\heatblue{23}{36.41} & \heatblue{29}{17.15} &
5.44 & 5.02 &
\heatorange{58}{50.60} & \heatorange{55}{48.93} &
\heatorange{48}{69.56} & \heatorange{47}{68.23} \\

DiffAnon   & 0.2 & $\psi$ &
\heatblue{49}{43.91} & \heatblue{46}{22.40} &
\heatblue{45}{41.82} & \heatblue{42}{18.87} &
5.73 & 5.38 &
\heatorange{46}{48.56} & \heatorange{35}{46.46} &
\heatorange{36}{65.58} & \heatorange{34}{63.76} \\

DiffAnon   & \textit{null} & $\psi$ &
\heatblue{55}{45.03} & \heatblue{53}{23.82} &
\heatblue{47}{42.43} & \heatblue{55}{20.66} &
5.79 & 5.61 &
\heatorange{38}{47.38} & \heatorange{25}{45.23} &
\heatorange{32}{64.32} & \heatorange{30}{62.45} \\

DiffAnon   & \textit{null} & $\psi$, spk CFG ($w_{spk}=3)$ &
\heatblue{64}{48.44} & \heatblue{64}{27.37} &
\heatblue{64}{48.16} & \heatblue{64}{22.78} &
6.63 & 6.22 &
\heatorange{10}{42.74} & \heatorange{10}{43.39} &
\heatorange{10}{57.05} & \heatorange{10}{56.06} \\
\hline
DiffAnon   & \textit{null} & \textit{null} &
\heatblue{54}{44.86} & \heatblue{46}{22.49} &
\heatblue{44}{41.52} & \heatblue{53}{20.46} &
5.92 & 5.61 &
\heatorange{36}{47.03} & \heatorange{32}{46.08} &
\heatorange{32}{64.09} & \heatorange{30}{62.56} \\

DiffAnon \textit{+ pitch-shift}  & 1.0 & $\psi$ &
\heatblue{27}{39.02} & \heatblue{17}{16.47} &
\heatblue{23}{36.43} & \heatblue{11}{14.72} &
5.21 & 4.88 &
\heatorange{70}{52.57} & \heatorange{58}{49.36} &
\heatorange{63}{74.36} & \heatorange{63}{73.17} \\
\bottomrule
\end{tabular}
\label{tab:results}
\caption*{\footnotesize * denotes reported results. Utility (UAR) evaluation is performed on IEMOCAP; other evaluations use Librispeech.}
\vspace{-10mm}
\end{table*}

Here, $\emptyset$ denotes a null condition, which means adding $0$ in conditioning. The base prediction uses only content and pseudo-speaker conditions. The guidance weight $w_{pro}$ interpolates between prosody-unconditional and prosody-conditional generations, providing explicit and continuous control over the utility–privacy trade-off through prosody preservation. We always use content condition $c_{sem}$ to ensure content preservation and pseudo-speaker condition $\psi$ to ensure guiding away from the source speaker. We experimented with multiple conditional–unconditional combinations, and this formulation proved most effective for isolating and controlling source prosody.

\textbf{Pseudo-speaker guidance.} In addition to prosody-adjusted generation, we explore pseudo-speaker guidance. In this mode, the prosody condition is always set to $\emptyset$, and privacy is enhanced by strengthening the pseudo-speaker conditioning. This formulation is given as:
\setlength{\abovedisplayskip}{3pt}
\setlength{\belowdisplayskip}{3pt}
\begin{equation}
\begin{aligned}
 \hat{\mathbf{x}}_{0,\theta}^{\mathrm{CFG}}
&=
(w_{spk} + 1 )  \hat{\mathbf{x}}_{0,\theta}(x_t,t,c_{sem}, \emptyset,\psi) \\
&\quad - w_{spk} \hat{\mathbf{x}}_{0,\theta}(x_t,t,c_{sem},\emptyset,\emptyset).
\end{aligned}
\end{equation}

Here, $w_{spk}$ controls the strength of the speaker-conditioned generation. As in standard CFG, subtracting the unconditional prediction sharpens the conditional signal, directing the output more strongly toward the pseudo-speaker.


To enable CFG at inference, the model is trained with randomly dropped conditions. Specifically, 50\% of the time it is trained with all conditions $\hat{\mathbf{x}}_{0,\theta}(x_t,t,c_{sem}, c_{pro},c_{spk})$, 30\% with prosody dropped $\hat{\mathbf{x}}_{0,\theta}(x_t,t,c_{sem}, \emptyset,c_{spk})$, and 20\% with both prosody and speaker dropped $\hat{\mathbf{x}}_{0,\theta}(x_t,t,c_{sem}, \emptyset,\emptyset)$. Most importantly, we avoid speaker-only dropping, as preliminary experiments showed that it encourages the model to exploit speaker information leaking through prosodic features, which is detrimental to anonymization.
 
\subsection{Model Architecture Details}
DiffAnon adopts a diffusion backbone inspired by NaturalSpeech2~\cite{shen2024naturalspeech}. The model comprises 40 WaveNet-style residual blocks~\cite{oord2016wavenetgenerativemodelraw}, each using a 1D non-dilated convolution with kernel size 5, 1024 channels. Timestep embeddings are projected via an MLP and added to the latent representations. Prosody and speaker conditions ($c_{pro}$, $c_{spk}$) are injected through separate non-dilated 1D convolutional projection layers before being added to the latent features, while the semantic condition ($c_{sem}$) is directly incorporated into each residual block.

\vspace{-2mm}
\section{Experimental Setup}

\textbf{Training \& Inference Details.} DiffAnon is trained on the training subsets of LibriTTS~\cite{libritts} for approximately 400k steps with a learning rate of $1\times10^{-4}$ and a batch size of 8 on a single NVIDIA H100 GPU. During inference, we use DDIM sampling~\cite{ddim} with 100 denoising steps. Pseudo-speakers are sampled from a pool constructed from LibriTTS training speakers, where each pseudo-speaker embedding is obtained by averaging the speaker embeddings across that speaker’s utterances.

\textbf{Evaluation.} We follow the official VoicePrivacy Challenge 2024 protocol~\cite{tomashenko2026voiceprivacychallengepreservingemotional}. Privacy is measured using equal error rate (EER)~\cite{KINNUNEN201012} under lazy-informed and semi-informed attacker scenarios. Utility is evaluated in terms of content preservation using word and character error rates (WER/CER) computed with an ASR system~\cite{speechbrain_v1}, and emotion preservation using speech emotion recognition (SER) accuracy (UAR)~\cite{pepino21_interspeech}. To further assess prosody preservation, we additionally report the official VoicePrivacy 2022 evaluation~\cite{vpc2022} on the \textit{libri-dev} and \textit{libri-test} subsets, including rank-based F0 correlation between source and anonymized speech. For comparison, we include the official baselines (B1–B6) and top-performing systems T8~\cite{henry}, T9~\cite{t9}, and T10~\cite{t10}.

\vspace{-3mm}
\section{Results}
We evaluate DiffAnon with prosody guidance weights
$w_{\mathrm{pro}} \in \{1, 0.8, 0.5, 0.2, 0\}$ to control source prosody preservation at inference. We also report pseudo-speaker guidance ($w_{\mathrm{spk}}=3.0$), an inference setting without CFG where both prosody and speaker conditions are set to null, and mean pitch-shifting applied before extracting prosody features. All results come from a single trained model, varying only inference-time settings.


\textbf{Controllable Prosody Preservation.} Table~\ref{tab:results} shows that DiffAnon enables continuous modulation of source prosody. With full prosody preservation ($w_{\mathrm{pro}}=1$), DiffAnon achieves the highest F0 correlation among its settings (76.67 on \textit{libri-dev}) and relatively strong emotion UAR (52.32), indicating effective retention of source prosodic structure. As $w_{\mathrm{pro}}$ decreases, both F0 correlation and UAR drop consistently. For example, F0 correlation decreases from 75.58 ($w_{\mathrm{pro}}=1$) to 69.56 ($w_{\mathrm{pro}}=0.5$) and further to 64.09 under null prosody, while emotion UAR decreases from 52.32 to 50.60 and 47.38, respectively. This trend coincides with improved privacy (e.g., \textit{libri-test} lazy EER increases from 33.09 to 42.43), confirming that CFG provides controlled prosody adjustment at inference and supports smooth interpolation between stronger anonymization and higher prosodic fidelity.

\textbf{Privacy.} Privacy results under lazy-informed and semi-informed attacker scenarios are reported in Table~\ref{tab:results}. Baselines span a wide range of privacy levels depending on architectural design, and DiffAnon similarly covers multiple privacy regimes via inference-time settings. The strongest privacy within DiffAnon is achieved under pseudo-speaker guidance with null prosody, reaching 48.16 EER on \textit{libri-test} (lazy), which is competitive with the strongest baseline systems. In semi-informed setting, the privacy decreases but remains competitive. As $w_{\mathrm{pro}}$ decreases, privacy improves consistently (e.g., from 33.09 to 42.43 EER (lazy) on \textit{libri-test}, and from 14.53 to 22.78 (semi)) demonstrating controllable adjustment of anonymization strength.

\textbf{Utility.} DiffAnon maintains stable and decent WER across inference settings, indicating effective content preservation through persistent semantic conditioning. On \textit{libri-test}, WER increases moderately from 4.62 ($w_{\mathrm{pro}}=1$) to 5.02 ($w_{\mathrm{pro}}=0.5$) and 5.61 under null prosody, remaining substantially lower than several privacy-focused baselines. Even under stronger pseudo-speaker guidance ($w_{\mathrm{spk}}=3$), WER remains reasonable at 6.22, showing robustness across operating points. For emotion preservation, DiffAnon achieves 52.32 UAR at full prosody preservation; although it is not explicitly designed for emotion, it outperforms most challenge baselines except systems optimized for emotion (e.g., T9 and T10). As $w_{\mathrm{pro}}$ decreases, UAR declines in a controlled manner (50.80 $\rightarrow$ 48.93 $\rightarrow$ 45.23), consistent with reduced prosodic fidelity, while privacy improves. Overall, DiffAnon preserves strong utility while enabling controllable anonymization. As CFG is a principled generation paradigm, the model can synthesize speech even under partially or fully dropped conditions, maintaining reasonable speech utility across all inference settings.

\begin{figure}[!t]
    \centering
    \scalebox{0.35}{\includegraphics{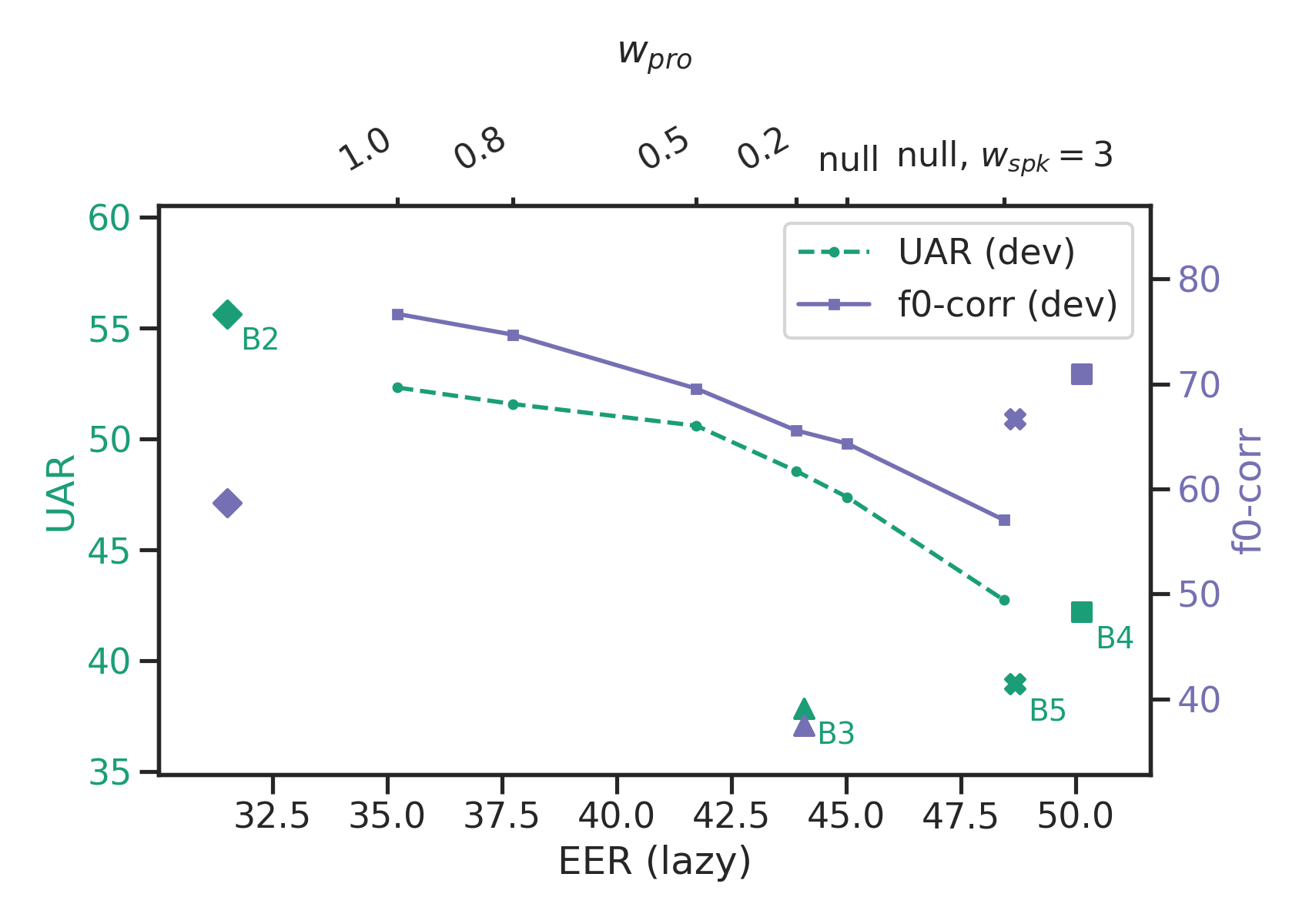}}
    \vspace{-6mm}
    \caption{DiffAnon's navigation in utility vs privacy trade-off curve (top axis represents different $w_{pro}$ settings)}
    \label{fig:tradeoff}
  \vspace{-5mm}
\end{figure}

\textbf{Privacy-Utility Tradeoff.} The privacy–utility trade-off is clearly observable within DiffAnon across inference settings (Table~\ref{tab:results}, Fig.~\ref{fig:tradeoff}). As $w_{\mathrm{pro}}$ decreases, privacy improves while prosodic fidelity declines in a structured manner. For example, increasing privacy from 33.09 to 42.43 (\textit{libri-test} lazy EER) corresponds to a decrease in F0 correlation from 75.58 to 62.45 and a reduction in emotion UAR from 50.80 to 45.23, in a monotonic way. In contrast to baselines that operate at fixed design points, DiffAnon provides explicit and continuous adjustment of the privacy–utility trade-off within a single trained model. 
%

\vspace{-2mm}
\section{Conclusion}
Prosody lies at the center of the utility–privacy trade-off in voice anonymization: preserving it crucial for expressiveness, but increases the risk of identity leakage. In this work, we introduced DiffAnon, a diffusion-based anonymization framework that enables explicit and continuous control over prosody preservation via classifier-free guidance.  A central contribution of this work is the empirical demonstration within a unified framework that prosody is a principal driver of the utility–privacy trade-off. By systematically modulating prosodic preservation, we show that changes in prosody directly and predictably alter both privacy and utility, establishing prosody as an important attribute in privacy-preserving speech systems. Future work will extend controllable anonymization to additional prosodic factors, such as duration, which also influence utility--privacy trade-off.




\section{Acknowledgments}
This work was supported by:
\begin{itemize}
    \item The National Science Foundation (NSF) CAREER Award IIS-2533652.
    \item The Office of the Director of National Intelligence (ODNI), Intelligence Advanced Research Projects Activity (IARPA), via the ARTS Program, Contract \#D2023-2308110001.
\end{itemize}
The views and conclusions contained herein are those of the authors and should not be interpreted as necessarily representing the official policies, either expressed or implied, of ODNI, IARPA, or the U.S. Government. The U.S. Government is authorized to reproduce and distribute reprints for governmental purposes notwithstanding any copyright annotation therein.

\section{Generative AI Use Disclosure}
Generative AI tools were employed solely for language polishing of text written by the authors. These tools were not used to generate scientific content, results, experimental designs, analyses, or conclusions. All authors are responsible for the full content of this paper and consent to its submission.

\bibliographystyle{IEEEtran}
\bibliography{mybib}

\end{document}